# Orbital order-disorder transition in $La_{1-x}Nd_xMnO_3$ (x = 0.0-1.0) and $La_{1-x-y}Nd_xSr_yMnO_3$ (x = 0.1, y = 0.05, 0.1)

Dipten Bhattacharya, P.Sujatha Devi, and H.S. Maiti
*Electroceramics Division, Central Glass and Ceramic Research Institute, Calcutta 700 032, India*

The nature of orbital order-disorder transition has been studied in the $La_{1-x}Nd_xMnO_3$ (x = 0.0-1.0) series which covers the entire range between two end points – $LaMnO_3$ and $NdMnO_3$ – as well as in $La_{0.85}Nd_{0.1}Sr_{0.05}MnO_3$ and $La_{0.8}Nd_{0.1}Sr_{0.1}MnO_3$. It has been observed that the first order nature of the transition gives way to higher order with the increase in ‘x’ in the case of pure manganites. The latent heat (L) associated with the transition, first, drops with a steeper slope within x = 0.0-0.3 and, then, gradually over a range $0.3<x<0.9$. This drop could, possibly, be due to evolution of finer orbital domain structure with ‘x’. In the case of Sr-doped samples, the transition appears to be of higher order nature even for a doping level 5 at%. In both cases, of course, the transition temperature $T_{JT}$ rises systematically with the drop in average A-site radius $\langle r_A \rangle$ or rise in average Mn-O-Mn bond bending angle <cos$^2$ö> while no apparent correlation could be observed with doping induced disorder ó$^2$. The cooperative nature of the orbital order, therefore, appears to be robust.

The issue of orbital order-disorder transition in pure and doped manganites has attracted a great deal of attention in recent times.[1] It has been widely noted that there is a close interplay among charge, spin, and orbital degrees of freedom in rare-earth perovskite manganites which gives rise to inter-dependence among orders in different degrees of freedom. Recent studies[2] have shown that the lattice distortion plays an important role in governing the interplay. More specifically, the interplay between cooperative Jahn-Teller (JT) distortion and $GdFeO_3$-type distortion in several transition metal oxide systems leads to a change in the JT distortion from a- to d-type.[3] The nature of the static orbital order, therefore, depends strongly on such interplay and can vary from conventional antiferro C-type to CE-type and even to a geometrically frustrated type for $90^o$ metal-oxygen-metal bonds.[4] The nature of the order-disorder transition too, appears to be dependent on coupling between JT order and lattice or interplay between JT and $GdFeO_3$-type distortion. Recent works[5,6] have pointed out that while the transition in the case of pure $LaMnO_3$ is first order in nature it becomes second order in the case of $PrMnO_3$ or $NdMnO_3$ where average A-site radius $\langle r_A \rangle$ is smaller. In the case of Ba-doped systems, the transition becomes second order beyond a very low doping level (2.5 at%).[7] The reason behind such cross-over in the nature of transition is not quite clear at this moment. In the case of Ba-doped systems, the higher order transition is proposed to be due to drop in anharmonic coupling between JT distortion and volume strain.[8] Strong coupling results in a large volume contraction at the transition, as observed in $LaMnO_3$, and hence a first order transition. Weak coupling, on the other hand, results in a higher order transition. In the case of single-valent systems, enhanced Mn-O-Mn bond bending due to smaller ion at A-site, possibly, results in a finer orbital domain structure. Such a system undergoes a broader, reminiscent of higher order, transition near $T_{JT}$.

Given such rich background, it is pertinent to raise few important points which have, to the best of our knowledge, not been addressed so far in the published literature: (i) how and when the orbital order-disorder transition turns higher order as a function of rise in <cos$^2$ö> (which quantifies the $180^o$-ö bending angle of Mn-O-Mn bonds) or drop in $\langle r_A \rangle$; (ii) whether the cross-over from first to second order transition follows similar patterns in both the cases: with drop in $\langle r_A \rangle$ in pure manganites and with rise in "Sr" doping level in doped ones; (iii) nature of the transition and drop in transition temperature ($T_{JT}$) with Sr-doping level in a smaller $<r_A>$ system; (iv) whether doping induced disorder has any role to play in governing $T_{JT}$. In this paper, we address these issues by studying the orbital order-disorder transition in the series $La_{1-x}Nd_xMnO_3$ (x = 0.0-1.0) which covers the entire range between $LaMnO_3$ and $NdMnO_3$. In addition, we have also studied the role of Mn-O-Mn bond bending in governing the drop in $T_{JT}$ with Sr-doping level. We have employed resistivity, differential thermal analysis (DTA) and differential scanning calorimetry (DSC) to observe the transition and its nature in all such systems. We found that, although, the latent heat of transition does drop with Nd-substitution level (x) as expected, the pattern of drop varies with 'x': it drops sharply within x = 0.0-0.3 and, then, gradually to zero across the $0.3<x<0.9$. On the other hand, Sr-doping gives rise to higher order transition even below a doping level 5-10 at%. The $T_{JT}$, in the case of pure manganites, rises almost linearly with 'x' having no apparent dependence on variance $\sigma^2$ while its drop, in the case of Sr-doped systems, appears to be influenced by bond bending due to smaller average A-site radius $<r_A>$. In both the cases, there is no apparent signature of phase segregation which points out the robustness of the cooperative nature of orbital order network.

The experiments have been carried out on high quality single phase bulk polycrystalline samples which have been prepared by using the powder obtained from solution chemistry technique – autoignition of a gel formed following boiling of mixed aqueous metal nitrate solution in presence of citric acid. The heat treatment has been carried out in argon atmosphere for maintaining the oxygen stoichiometry. The actual $Mn^{4+}$ concentration has been estimated by redox titration technique[9] and is found to vary between 2-10% in the samples belonging to $La_{1-x}Nd_xMnO_3$ series. The samples have been further characterized by scanning electron microscopy (SEM), energy dispersive X-ray spectroscopy (EDX), and X-ray diffraction (XRD) study. The four-probe resistivity has been measured across a temperature range 300-1200 K under vacuum ($\sim10^{-3}$ torr) while DTA (Shimadzu, DTA-50) and DSC (Pyris Diamond DSC, Perkin-Elmer) studies have been carried out under inert atmosphere (nitrogen) across the temperature ranges 300-1273 K and 300-973 K, respectively. High quality platinum paste is used for making the contacts for resistivity measurement.

In Fig. 1, we show the lattice parameters and lattice volume, estimated from room temperature XRD, for the undoped samples. For all the samples, the room temperature phase appears to be orbital-ordered orthorhombic O' with $c/\sqrt{2}<a<b$ (space group Pbnm). The lattice volume collapses with the increase in 'Nd' substitution, as expected. The collapse is sharp within x = 0.0-0.3 and gradual beyond that. This pattern appears to be similar to the pattern of drop in latent heat and could possibly have a correlation. The room temperature orthorhombic lattice distortion D increases systematically with the drop in average A-site radius $\langle r_A \rangle$. We have also estimated the disorder $\sigma^2$ using the relation $\sigma^2 = \sum_{i=1}^{n} x_i (r_i - \langle r_A \rangle)^2$. These parameters are listed in Table-I for all the compositions.

In Fig. 2, we show the resistivity vs. temperature curves for the pure manganites. A reasonably sharp drop in resistivity is observed near $T_{JT}$ for the samples with Nd-substitution level x<0.5. The sharp drop highlights the quality of the samples used for the study. The nature of the transition appears to be broader beyond x = 0.5, yet $T_{JT}$ could be identified clearly for all the samples. In the first inset to the Fig. 2, we show the temperatures $T^*$ (below which static orbital order is present) and $T_{JT}$ (beyond which orbital fluctuations are present).[5] In the second inset, we plot the $T_{JT}$ vs. x which reflects a remarkable linear dependence of $T_{JT}$ on x. The $T_{JT}$ does not seem to be influenced by disorder $\sigma^2$. Nearly linear rise in $T_{JT}$ with 'x' rules out the possibility of presence of impure phases or phase segregation. It is noteworthy here that in spite of the presence of finite amount of $Mn^{4+}$ in the samples belonging to $La_{1-x}Nd_xMnO_3$ series, such a linear rise in $T_{JT}$ could be observed. This observation highlights that the influence of substitution of a smaller ion (like 'Nd') at 'La'-site can be clearly observed even in presence of finite amount of $Mn^{4+}$. From the resistivity patterns below $T^*$ and above $T_{JT}$, we estimate the activation energies ($E_a$) corresponding to the appropriate temperature ranges using the thermally activated hopping transport relation $\rho = \rho_0 T.\exp(E_a/k_BT)$. The values of $E_a$ for all the samples are listed in Table-I. There is an overall trend of increase in $E_a$ with ‘x’ which tallies with the pattern reported in Ref. 5. In one or two cases, however, the resistivity as well as $E_a$ appear to be anomalous which could be due to variation in sintering and hence the microstructure of the samples. In Fig. 3, we plot the latent heat associated with the order-disorder transition as a function of ‘x’. In the inset, we show representative DTA and DSC thermograms. The thermograms depict endothermic peaks for all the samples right up to x = 0.7. We estimated the change in enthalpy or latent heat from the area under the peak, after proper background suppression. As expected, the latent heat (L) depends on ‘x’; but, interestingly, the nature of drop varies with ‘x’. The slope of the ‘L’ vs. ‘x’ plot changes at x = 0.3: from ~ 20/30 J/gm/at% to ~5/40 J/gm/at%. Therefore, it appears that the cross-over, from first to higher order transition in presence of lattice distortion in the $La_{1-x}Nd_xMnO_3$ series, is following two distinct sequences: a sharp drop and a gradual change. This pattern, probably, indicates that as ‘La’ is substituted by ‘Nd’, the domain size of the static orbital order network decreases sharply within x = 0.0-0.3. Beyond that level, the change is more gradual. It is noteworthy that the overall pattern of ‘L’ vs. ‘x’ does not exhibit any anomaly near x = 0.5 where the disorder $\sigma^2$ is maximum. Rather, the systematic drop in ‘L’ points out overpowering effect of increased lattice distortion due to substitution of ‘La’ by smaller ion ‘Nd’. The ‘L’ vs. ‘x’ pattern tallies roughly with the pattern of variation of lattice volume shown in Fig. 1 and helps in tracking down the exact nature of the cross-over as a function of $<r_A>$ (or $<\cos^2\phi>$). This is *one of the central results of this paper*.

We now turn to the Sr-doped samples. In this case, following interesting observations have been made: (i) for both the $La_{0.85}Nd_{0.1}Sr_{0.05}MnO_3$ and $La_{0.8}Nd_{0.1}Sr_{0.1}MnO_3$ samples, we observe single order-disorder transition temperature ($T_{JT}$) in spite of large $\sigma^2$ which points out that the cooperative structure of the orbital order network is retained; (ii) the drop in $T_{JT}$ with Sr-doping level is influenced by Mn-O-Mn bond bending due to smaller ions (like Nd) by nearly the same extent as in undoped series; therefore, the $T_{JT}$-$<r_A>$ or $T_{JT}$-bond bending relation holds good even in presence of “Sr”; the $Mn^{4+}$ concentration in these cases is found to be within 14-17% and thus, close to the boundary of insulator-metal transition; (iii) the order-disorder transition appears to be of higher order nature. In Fig. 4, we show the resistivity vs. temperature patterns. $T_{JT}$ could be clearly identified from the resistivity pattern. However, no peak could be observed in DTA/DSC patterns which signals higher order transition at $T_{JT}$. In the inset of Fig. 4, we show the drop in $T_{JT}$ with Sr-doping level in presence and absence of Nd at A-site. This particular result demonstrates that the drop in $T_{JT}$ due to rise in "Sr" can be, at least, partially compensated by reducing the $\langle r_A \rangle$. This has also been observed, previously, when the drop in $T_{JT}$ with doping level 'x' has been studied in $La_{1-x}Sr_xMnO_3$ and $La_{1-x}Ca_xMnO_3$; the pattern turns out to be broader in the latter case.[10] These results again point out that the cooperative structure of the orbital order network is robust and does not collapse because of high $\sigma^2$. However, one significant observation is the cross-over in the nature of transition. In the case of Sr-doped samples, the cross-over appears to be taking place within even smaller doping regime (<5 at%). Earlier, it has been shown[7] for the Ba-doped samples that the cross-over is taking place beyond 2.5 at%. This is, certainly, in contrast to the observation made in the case of undoped manganite series and *is another important result of this paper*.

The reason behind the cross-over from first to higher order transition is still not well established. Recently, it has been proposed[8] that the anharmonic coupling between JT distortion and volume strain has a role to play in determining the volume collapse or nature of the transition near $T_{JT}$. Below a certain value of the coupling parameter, the nature of transition changes from first to higher. On the other hand, local measurements of the orbital order network using coherent X-ray beam[11] indicates presence of orbital domain structure and reduction in the domain size with the increase in lattice distortion - from 4000-7000 Å in $LaMnO_3$ to roughly 320 Å in $Pr_{0.6}Ca_{0.4}MnO_3$. It has also been pointed out that with the increase in Mn-O-Mn bond bending a frustrated structure evolves.[4] In a recent report, observation of frustration of $e_g^1$ level due to smaller ion at A-site and consequent drop in Neel point $T_N$ has been presented.[12] The mechanism of drop in coupling between JT distortion and volume strain could be useful in describing the cross-over in 'Ba' or 'Sr' doped systems where one observes a rise in orbitally disordered metallic ferromagnetic phase. However, it is not immediately clear whether it holds good for the pure manganites where lattice distortion systematically increases and static orbital order is found to prevail. In fact, in "Sr"-doped systems, possibility of "orbital liquid" phase formation due to quantum fluctuations has been highlighted.[13] It should also be noted, in this context, that the time resolved study in Ref. 11 exhibits fluctuation in the orbital domain structure in the case of $Pr_{0.6}Ca_{0.4}MnO_3$ system within the laboratory time scale while pure $LaMnO_3$ exhibits a static order.

Therefore, it is quite possible that in our case with the increase in lattice distortion due to 'Nd' substitution, orbital domain size drops and eventually either a fine domain structure or a frustrated structure evolves. The volume fraction of long-range static orbital order network comes down which results in the drop in latent heat with 'x'. The transition at $T_{JT}$ in the case of conventional C-type orbital order is akin to solid-liquid melting into a fluctuating phase[14] while in the case of systems like $PrMnO_3$ or $NdMnO_3$, with static short range order, it is more like the transitions observed in relaxor or glassy system where the local fluctuations freeze at the transition point in the dynamic sense. The thermodynamic signature of the phase transition in such cases is controversial.[16] There is another possibility. In systems with local lattice distortions or defects, one can observe a local first order transition yet a global rounding of the transition due to wide distribution of the transition point.[17] With progressive substitution of 'La' by 'Nd', increased lattice distortion can interfere with cooperative JT order and give rise to evolution of finer domain structure with distributed $T_{JT}$. A local structural and calorimetric measurement with proper spatio-temporal resolution can solve this issue.[18]

In summary, we show that the pattern of cross-over in the nature of orbital order-disorder transition in single-valent $La_{1-x}Nd_xMnO_3$ series (x = 0.0-1.0) depends systematically on 'x'. In contrast, in the case of Sr-doping, the cross-over is rather fast and takes place even below a doping level 5 at%. In presence of smaller ion at A-site, the transition temperature does not drop as sharply with the Sr-doping level as observed in the absence of smaller ion. The transition temperature does not depict any clear correlation with the doping induced disorder for both the pure and Sr-doped systems. All these reflect the robustness of the cooperative structure of the orbital order network. The reason behind cross-over in the nature of orbital order-disorder transition in single-valent systems could possibly be development of fine domain structure or a glassy structure with short range order. Local measurements with proper spatio-temporal resolution can settle this issue.

We thank A. Kumar for assistance in chemical analysis and P. Choudhury for helpful discussion. This work is carried out under the CSIR networked program "custom-tailored special materials" (CMM 0022).

Table I. List of few relevant parameters of the single-valent and doped manganites; Orthorhombic distortion $D = \Sigma|a_i\text{-}a|/3a_i$; $a = (abc/\sqrt{2})^{1/3}$, $a_1 = a$, $a_2 = b$, $a_3 = c/\sqrt{2}$

| Compositions | $<r_A>$ (Å) | D (%) | ó$^2$ ($10^{-6}$ Å$^2$) | $E_a$ (eV) Below $T^*$ | $E_a$ (eV) Above $T_{JT}$ |
|---|---|---|---|---|---|
| $LaMnO_3$ | 1.216 | 1.84 | 0 | 0.22 | 0.35 |
| $La_{0.95}Nd_{0.05}MnO_3$ | 1.2134 | 1.974 | 133 | 0.21 | 0.33 |
| $La_{0.9}Nd_{0.1}MnO_3$ | 1.2107 | 2.076 | 252 | 0.22 | 0.31 |
| $La_{0.8}Nd_{0.2}MnO_3$ | 1.2054 | 2.256 | 449 | 0.198 | 0.14 |
| $La_{0.7}Nd_{0.3}MnO_3$ | 1.2001 | 2.524 | 590 | 0.199 | 0.42 |
| $La_{0.5}Nd_{0.5}MnO_3$ | 1.1895 | 2.66 | 702 | 0.24 | 0.85 |
| $La_{0.4}Nd_{0.6}MnO_3$ | 1.1842 | 2.74 | 674 | 0.25 | 1.16 |
| $La_{0.3}Nd_{0.7}MnO_3$ | 1.1789 | 2.84 | 590 | 0.24 | 0.98 |
| $La_{0.1}Nd_{0.9}MnO_3$ | 1.1683 | 3.645 | 252 | 0.27 | 1.03 |
| $NdMnO_3$ | 1.163 | 3.748 | 0 | 0.26 | 1.63 |
| $La_{0.85}Nd_{0.1}Sr_{0.05}MnO_3$ | 1.2154 | 1.23 | 722 | 0.26 | 0.2 |
| $La_{0.8}Nd_{0.1}Sr_{0.1}MnO_3$ | 1.2201 | 0.95 | 1147 | 0.16 | 0.19 |

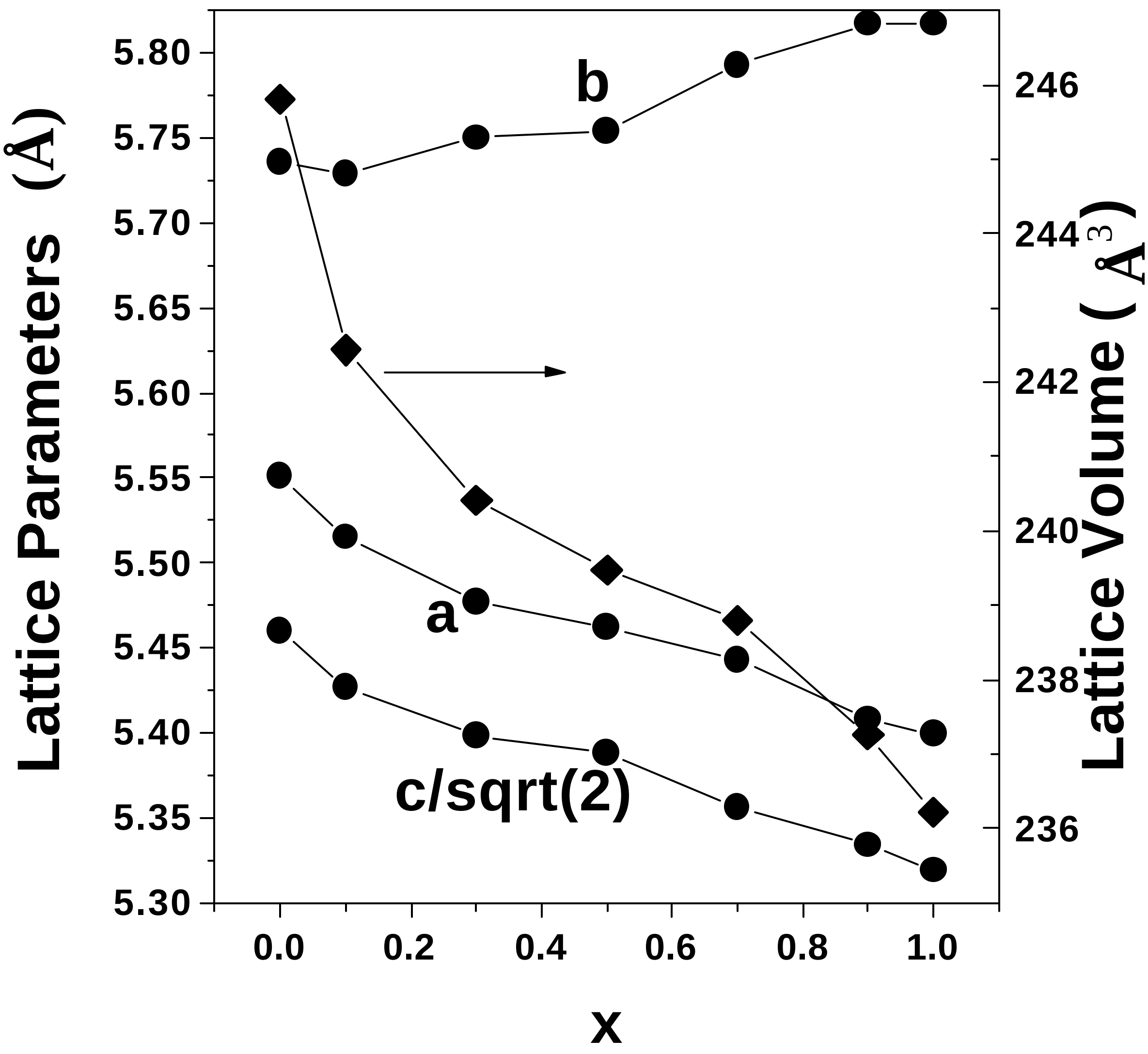

Fig.1. Variation of the lattice parameters and lattice volume with the 'Nd' substitution level 'x' in $La_{1-x}Nd_xMnO_3$ series as estimated from the room temperature X-ray diffraction pattern

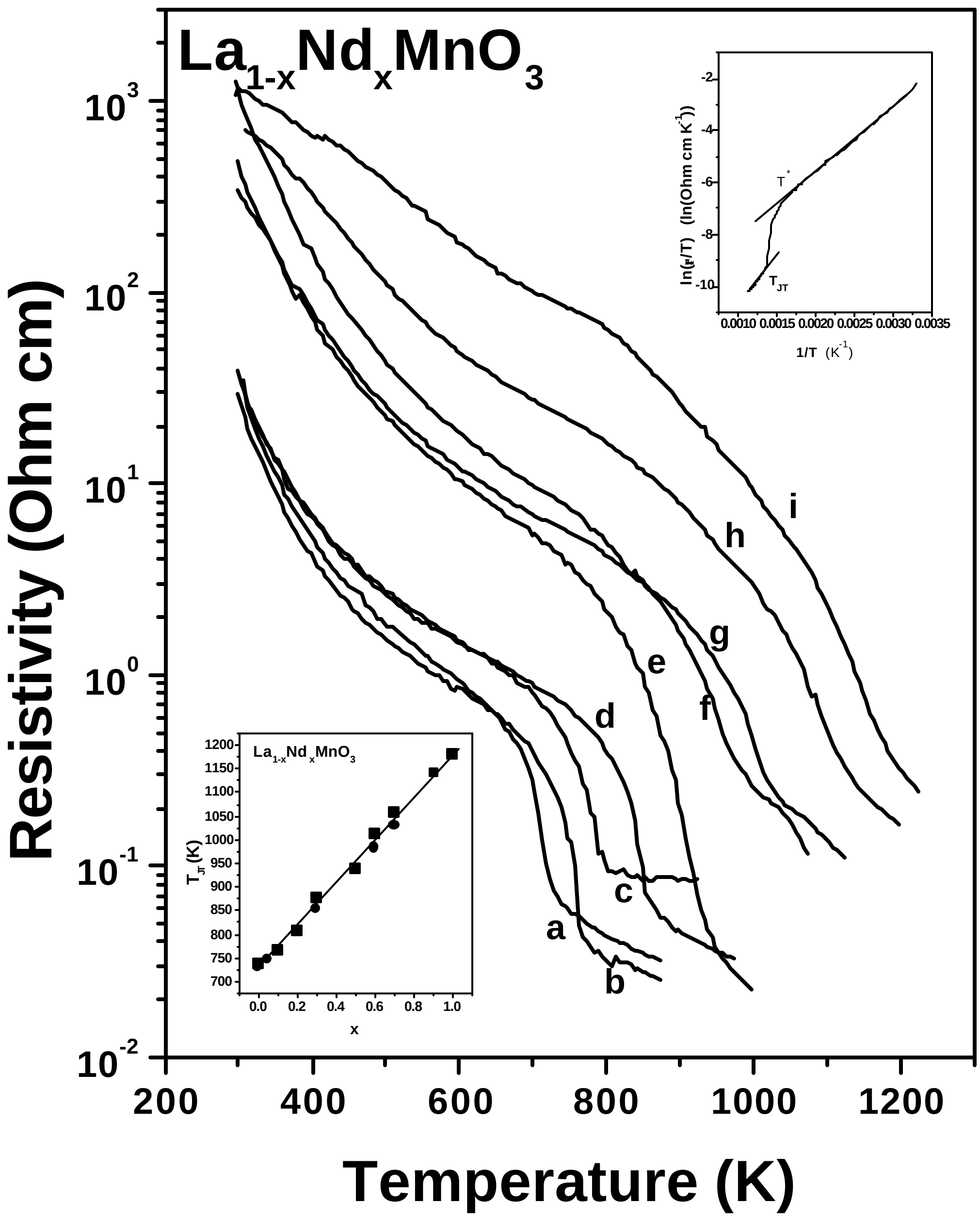

Fig.2. Resistivity vs. temperature patterns for the $La_{1-x}Nd_xMnO_3$ series; a: x = 0.0, b: x = 0.1, c: x = 0.2, d: x = 0.3, e: x = 0.5, f: x = 0.6, g: x= 0.7, h: x = 0.9, i: x = 1.0. Insets: (top) the temperatures $T^*$ (below which static order prevails) and $T_{JT}$ (above which fluctuating Jahn-Teller order dominates) in $LaMnO_3$, for example; (bottom) $T_{JT}$ vs. 'x' as observed in resistivity (square symbol) and DTA/DSC (circle symbol) measurements.

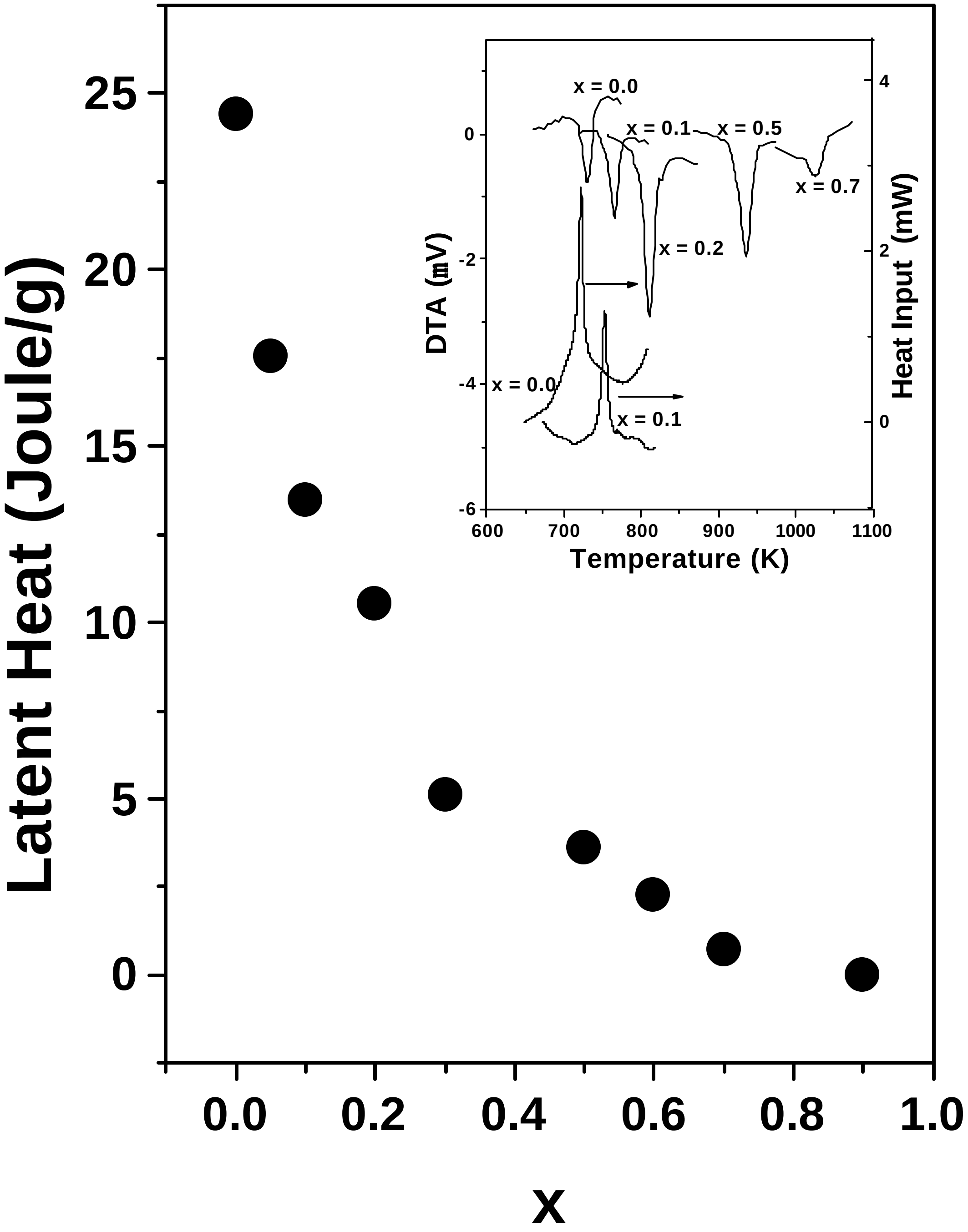

Fig.3. Change in enthalpy at $T_{JT}$, as observed in DTA/DSC measurements, as a function of 'Nd' substitution in $La_{1-x}Nd_xMnO_3$ series; inset: typical DTA/DSC thermograms. It is to be noted that the mass of each sample was different.

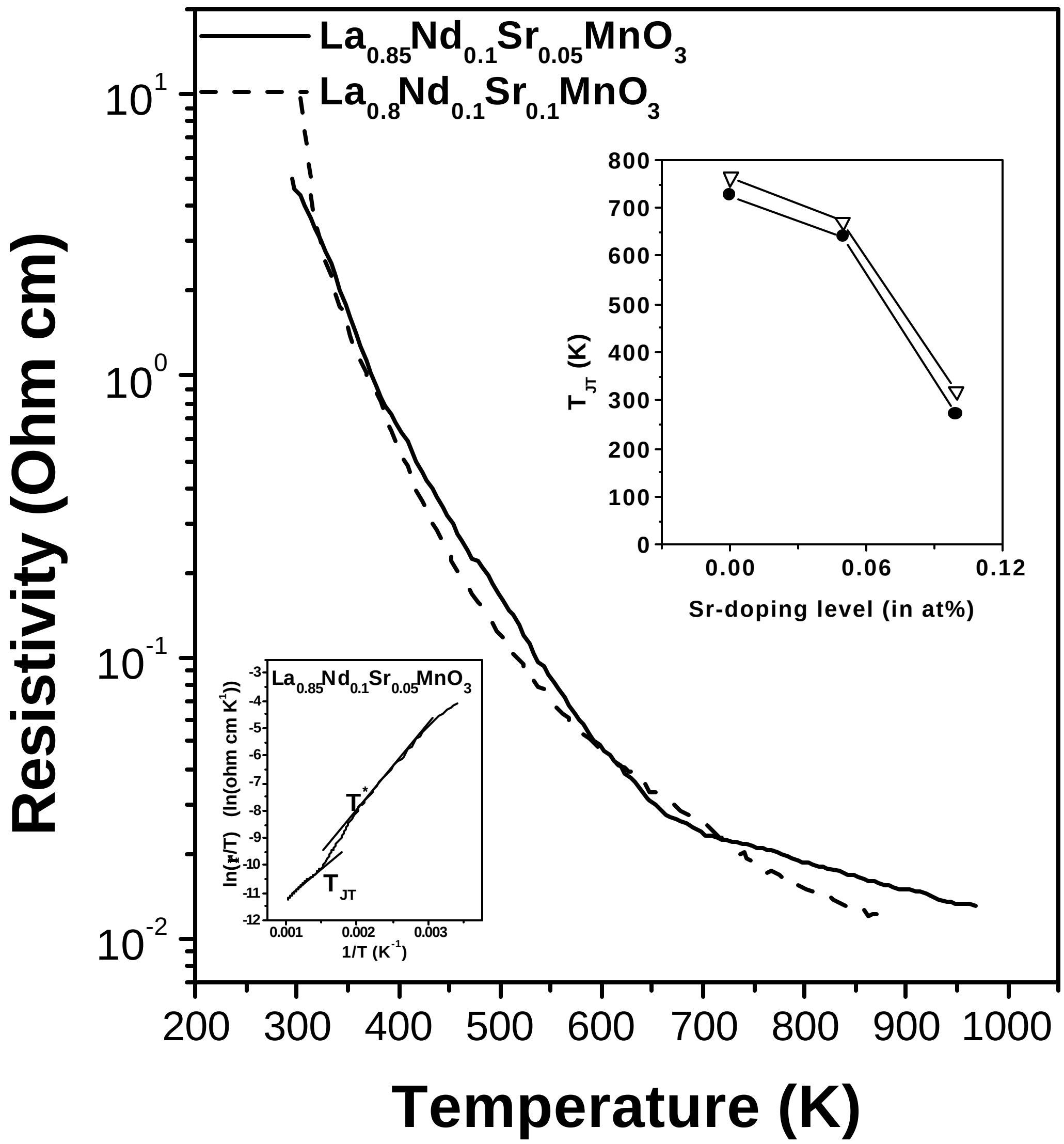

Fig. 4. Resistivity vs. temperature patterns for the Sr-doped La-Nd systems; insets: (top) $T_{JT}$ vs. Sr-doping level for pure La-Sr (solid symbol) and La-Sr-Nd (open symbol) systems; the difference in $T_{JT}$ due to 'Nd' doping is palpable; (bottom) ln(ρ/T) vs. 1/T plot for $La_{0.85}Nd_{0.1}Sr_{0.05}MnO_3$, for example.